\documentclass[reprint, superscriptaddress, amsmath,amssymb, aps, prx, longbibliography,]{revtex4-1}
\usepackage{braket}
\usepackage{graphicx}
\usepackage{amsmath}
\usepackage{amssymb}
\usepackage{algorithm}
\usepackage{algpseudocode}
\usepackage{bbm}
\usepackage[colorlinks=true,urlcolor=blue,linkcolor=blue,citecolor=blue]{hyperref}
\usepackage[margin=0.75in]{geometry}

\usepackage{times}

\begin{document}

\title{On the Notion of Dark Space-Time and Quantum Entanglement}

\author{Arun K. Pati}

\email{patiqubit@gmail.com}
\affiliation{Synergy Quantum Pvt Ltd, Second Floor, Research and Innovation Park\\ 
Indian Institute of Technology Delhi, Hauz Khas, New Delhi, India}

\date{\today}

\begin{abstract}
The nature of quantum nonlocality, as exemplified by entanglement, remains one of the deepest mysteries in quantum physics, challenging classical notions of causality and locality. In this work, we introduce the concept of dark space-time, a hidden geometric structure that coexists with ordinary space-time but may allow superluminal information transfer. We propose a modified space-time metric for dark space-time, in which the speed of causal influences exceeds the speed of light, thereby permitting nonlocal correlations to be naturally mediated through an unobservable channel. The framework is developed through a two-space-time quantum formalism, where entangled states evolve via interactions between ordinary and dark space-time sectors.  Furthermore, we discuss the implications of  dark space-time for ER=EPR conjecture and black hole information paradox. The notion of dark space-time is more fundamental than the dark energy and dark matter. 
Our results provide a novel approach to reconciling quantum mechanics with a deeper geometric foundation, offering insights into the fundamental nature of reality.
\end{abstract}

\maketitle

\section{Introduction}
Modern physics is built upon two foundational frameworks: general relativity (GR) and quantum mechanics (QM). General relativity provides a geometric understanding of space-time and gravity, while quantum mechanics governs the behavior of matter and energy at microscopic scales. However, despite their individual successes, these two theories remain incompatible at a fundamental level. Attempts to unify them into a consistent framework of quantum gravity have led to deep conceptual puzzles and paradoxes \cite{cow}.

One of the most intriguing challenges in this unification is the nature of space-time itself at the quantum level. In classical general relativity, space-time is a continuous, background fabric. In quantum mechanics, however, space-time should exhibit quantum properties—yet no direct quantum description of space-time exists in current physics. Instead, space-time is often treated classically, leading to contradictions in extreme scenarios such as quantum nonlocality and black hole information paradoxes.

On the other hand, astrophysical observations have revealed that the vast majority of the universe consists of dark matter and dark energy, neither of which can be explained within the Standard Model of particle physics \cite{miao}. This raises a profound question:
Could space-time itself have a "dark" counterpart, just as matter and energy do? If so, dark space-time could provide a missing link between quantum mechanics and gravity. Thus, dark space-time  would represent an additional geometric structure beyond ordinary space-time, influencing quantum evolution in a subtle but significant way. This concept could explain: (i) Quantum nonlocality without violating causality, (ii) The origin of quantum entanglement as an emergent phenomenon, (iii) Possible Lorentz symmetry violations at high energies, (iv) New sources of decoherence, affecting quantum technologies and many more.

One of the strongest motivations for dark space-time comes from quantum entanglement. In quantum mechanics, two entangled particles remain strongly correlated, even across vast distances—a phenomenon Einstein famously called ``spooky action at a distance '' \cite{epr}. While standard quantum mechanics describes this using nonlocal correlations \cite{bell,chsh}, there is no underlying physical mechanism explaining how information is maintained without direct communication. If dark space-time is an intrinsic part of quantum reality, it could serve as an information-preserving structure beyond ordinary space-time. Instead of faster-than-light signaling, entangled states may be connected through an unseen dark geometric fabric, providing a deeper explanation for quantum nonlocality while respecting relativistic causality.

The introduction of dark space-time (DST) necessitates a fundamental reformulation of quantum mechanics, moving beyond its traditional description in a single space-time manifold. In standard quantum theory, quantum states evolve according to the Schr\"odinger equation (or its relativistic counterparts) within ordinary space-time, with no additional geometric structure. However, if DST exists as an independent but interacting geometric sector, then quantum states should be described not just in terms of ordinary space-time, but within a larger two-space-time system where both the visible and dark space-time sectors influence quantum evolution.

In this work, we propose the concept of dark space-time—a hidden geometric structure that exists alongside ordinary space-time and may facilitate superluminal information transfer. We introduce a modified space-time metric where causal influences can propagate faster than light, enabling nonlocal correlations to emerge through an undetectable yet causal channel. This framework is formulated within a two-space-time quantum formalism, in which entangled states evolve through interactions between ordinary and dark space-time.
Additionally, we explore the broader implications of dark space-time for fundamental physics, particularly its potential role in the ER=EPR conjecture \cite{rosen,mald} and the black hole information paradox. We argue that dark space-time represents a more fundamental structure than dark matter and dark energy, offering a unifying perspective on these cosmic mysteries. By embedding quantum mechanics within an extended geometric foundation, our approach provides new insights into the underlying nature of reality and the deep connection between space-time and quantum information.

\section{Dark space-Time Geometry}
In standard relativity, space-time is described by the Lorentzian metric.
However, if dark space-time is a distinct geometric structure with different causal properties, it may require a non-Lorentzian metric. Below, we propose possible non-Lorentzian metric structures for dark space-time. We leave the option open and only future experiemntal evidence can decide which metric is most plausible one.\\

\noindent
1. {\it Non-Lorentzian Metric for DST}:  A pseudo-Riemannian metrics, allowing space-time anisotropy. A possible candidate for DST is given by

\begin{eqnarray}
ds^2 = -c_{dark}^2 f(x,t) dt^2 + g_{ij}(x,t)  dx^i dx^j,
\end{eqnarray}
where $c_{dark} >> c$ is the effective speed of propgation in dark space-time, 
$f(x,t)$ is a space-time dependent function allowing for variable causal structure and 
$g_{ij}(x,t) $ represents the metric structure for spatial geometry in dark space-time.
If $f(x,t)$ is negative in certain regions, time ordering can change, allowing entanglement correlations to travel differently in
dark space-time than in ordinary space-time.\\

\noindent
2.{\it  Emergent Lorentz Violating Preferred Frame in DST}:
If dark space-time breaks Lorentz invariance, we can write an explicit preferred-frame metric as given by
\begin{eqnarray}
ds^2 = -(c^2 + \xi v^2)  dt^2 +  dx^2 + dy^2 + dz^2 ,
\end{eqnarray}
where $v$ is the velocity of an observer relative to a hidden DST frame, and  $\xi$ 
determines the strength of Lorentz violation. If  $\xi > 0$, then DST interactions depend on motion relative to a preferred frame, violating Special Relativity but possibly explaining superluminal correlations.\\

\noindent
3. {\it Hybrid Ordinary-Dark Space-Time}:
A possible two-metric model incorporates interactions between ordinary and dark space-time can be given by
\begin{eqnarray}
ds^2 = g_{\mu \nu}^{ordinary} dX_{\mu} dX^{\nu} + \lambda h_{\mu \nu}^{dark} dx_{\mu} dx^{\nu},
\end{eqnarray}
where $g_{\mu \nu}^{ordinary}$ is the standard Minkowski metric with $X_{\mu} $ as the ordinary space-time coordinate, 
$h_{\mu \nu}^{dark}$ is the metric in dark space-time with $x_{\mu} $ as the dark space-time coordinate,
and $\lambda$ is a coupling parameter determining the strength of interactions between ordinary and dark space-time.
This framework may allow for locality in dark space-time but apparent nonlocality in ordinary space-time.

\section{Dark space-time and Quantum Entanglement}
In this section, we will develop how the idea of dark space-time can explain the mystery of quantum entanglement. If dark matter and dark energy are hypothesized to interact with ordinary matter primarily through gravity while remaining undetectable electromagnetically, one could speculate that dark space-time represents an extension of general relativity where a hidden geometric structure underlies the fabric of the cosmos. Below, we list some possible properties of dark space-time:

\noindent
(i) Just as dark energy affects cosmic expansion, dark space-time could be an additional component modifying the space-time metric at small and large scales.

\noindent
(ii) Dark space-time could be modeled as an emergent quantum geometry that coexists with our known space-time but remains undetectable under standard measurements.

\noindent
(iii) Inspired by quantum gravity approaches like loop quantum gravity \cite{ashte} and string theory, dark space-time might exist as a superposition of multiple geometric states. Quantum coherence between dark and ordinary space-time could lead to nonlocal effects.

\noindent
(v) If dark space-time interacts weakly with ordinary space-time via gravity, it could be modeled as an entangled sector of the universe. 


\section{ Postulates of Dark Space-Time Structure}

The framework of dark space-time is based on a set of fundamental postulates that extend standard quantum mechanics and relativity to incorporate an additional hidden geometric structure. These postulates provide the foundation for describing the interaction between ordinary space-time and dark space-time, particularly in the context of quantum nonlocality and superluminal information transfer.\\

\noindent
{\bf Postulate 1. Existence of a hidden dark space-time:}
There exists a hidden dark space-time that coexists with ordinary space-time, forming a two-space-time structure. Unlike ordinary space-time, which obeys Lorentz symmetry and causality constraints, dark space-time allows for superluminal information transfer and nonlocal correlations.


The universe has ordinary space-time and dark space-time as a parallel manifold, i.e., it consists of two intertwined space-time manifolds
${\cal M} = {\cal M}_{ordinary} \times {\cal M}_{dark} $ and each manifold has its own metric tensor $g_{\mu \nu}$ and $h_{\mu \nu}$
These manifolds interact weakly via an entanglement-induced connection.\\

\noindent
{\bf Postulate 2. Quantum State in Two Space-Times:}
Quantum systems exist simultaneously in both ordinary and dark space-time, described by an extended wavefunction.
The universe is described by a quantum state $|\Psi \rangle  \in 
{\cal H} = {\cal H}_{ordinary} \otimes  {\cal H}_{dark} $. This wave function contains all of the information about the ordinary geometry, dark space-time geometry and matter content of the universe.\\

\noindent
{\bf Postulate 3. Quantum Evolution:}
The quantum state of the universe along with dark space-time satisfies a general Schr\"odinger equation
\begin{align}
   i\hbar \frac{d }{dt} |\Psi \rangle = H |\Psi \rangle,
\end{align}
where $H$ is the total Hamiltonian as given by 
$$H= H_{ordinary} + H_{dark} + H_{int}. $$
Here  $H_{ordinary}$ describes the Hamiltonian for standard space-time evolution, $H_{dark}$ governs dark space-time dynamics and 
$H_{int}$ represents the interaction between both sectors.\\

\noindent
{\bf Postulate 4. Dark Space-Time as a Hidden Variable Mediator:}
The apparent "instantaneous" collapse of entangled states is mediated by hidden interactions in the dark space-time, and quantum information (spooky action) propagates faster than light but remains inaccessible in direct measurements.

These postulates provide a coherent theoretical framework in which dark space-time acts as a hidden geometric structure mediating quantum nonlocality, explaining dark matter dynamics \cite{am}, and suggesting potential modifications to gravity and relativity. The study of
dark space-time could revolutionize our understanding of quantum mechanics, general relativity, and high-energy physics by unifying them under a deeper space-time structure.

In the dark space-time, we require extending the Hilbert space of quantum mechanics to account for hidden degrees of freedom associated with DST. If ${\cal H}_{O}$  be the Hilbert space of states evolving in ordinary space-time and ${\cal H}_{D}$  
be the Hilbert space corresponding to quantum states in dark space-time, then the total quantum state would then reside in a tensor product space ${\cal H}_{Total} = {\cal H}_{O} \otimes {\cal H}_{D}$. The dynamics of this extended system can be governed by a generalized Schrödinger equation or an effective quantum master equation obtained by tracing out the dark degrees of freedom. The interaction Hamiltonian between the two space-time sectors can introduces novel effects such as: (i) Decoherence due to information leakage into  
dark space-time, (ii) modifying standard quantum coherence times, (iii) 
Nonlocal correlations arising from dark space-time-mediated interactions, potentially explaining entanglement in a deeper geometric sense and (iv) 
Modified time evolution, leading to possible deviations from standard quantum mechanics, testable in high-precision quantum experiments.
This naturally leads to a quantum master equation incorporating hidden DST interactions, modifying standard quantum evolution in measurable ways. If DST acts as an additional information channel, quantum evolution may no longer be strictly unitary within ordinary space-time, instead acquiring corrections due to entanglement between visible and dark sectors. The effective density matrix evolution in ordinary space-time may thus take the form of a Lindblad-type equation, with DST contributions manifesting as additional dissipative or coherence-enhancing terms.

This reformulation of quantum mechanics in an extended space-time framework offers a powerful new paradigm—one that not only modifies fundamental quantum theory but also provides new insights into quantum gravity, high-energy physics, and the role of space-time in quantum information processing. For example, in black hole Information Paradoxes, one may argue that ff information is stored in DST, then black hole evaporation might not result in information loss. If dark matter moves in DST rather than ordinary space-time, its apparent gravitational behavior could be explained differently.
Furthermore, it may have implication in cosmic acceleration and dark energy. If DST has an intrinsic expansion rate different from ordinary space-time, it might naturally explain the observed acceleration of the universe.
This proposal may hint towards a new paradigm shift in our understanding.
The concept of dark space-time challenges our traditional view of reality, suggesting that quantum states may evolve in a hidden, extended geometric structure beyond classical space-time. Our work may lead to mathematical formalism, where one can explore whether DST provides a missing piece in our understanding of quantum mechanics, gravity, and cosmology.

Dark matter and dark energy have already revealed that our understanding of space-time is incomplete. Some natural motivations for dark space-time include: (i)Hidden Geometrical Degrees of Freedom: General relativity describes space-time using a four-dimensional Lorentzian manifold. Could there be an additional, hidden geometric structure that manifests as dark space-time?
(ii) Extra Dimensions: String theory and braneworld models suggest that additional spatial dimensions exist beyond the observed four. If these dimensions form a separate but coupled space-time, they may behave as "dark space-time."
(iii) Quantum Gravity Effects: In quantum gravity, space-time itself may emerge from entanglement (as suggested by ER=EPR). Could there exist an "entangled" but causally disconnected space-time that interacts indirectly with our own?
(iv)	Modified Gravity: The effects attributed to dark matter and dark energy might arise from modifications to space-time at large scales. This raises the question: Is there an additional "layer" of space-time contributing to these phenomenon?

If dark space-time exists, it could have the following novel properties: \\
(A)  Weak or Non-Local Gravitational Interaction:
(i) Dark space-time might interact with normal gravity weakly, explaining the anomalies in galactic rotation curves and cosmic expansion without invoking dark matter,
(ii) If the coupling constant between dark and visible space-time is tiny, effects would only appear on cosmological scales.

(B) Exotic Causal Structure: (i) If dark space-time has a different metric signature, time could flow differently or even not exist in the same way, (ii) It might explain faster-than-light information transfer in certain quantum processes (such as entanglement).

(C) Possible Explanation for Dark Energy: (i)	If dark space-time exerts a repulsive effect on our universe, it might naturally explain cosmic acceleration without invoking a cosmological constant.

(D) Influence on Quantum Mechanics: (i) If quantum fields extend into dark space-time, there could be hidden influences on quantum systems, (ii)	Dark space-time might provide a deeper explanation for wavefunction collapse or quantum nonlocality.

\section{Dark Space-Time and Non-Locality}
The proposal for dark space-time introduces a hidden, interacting quantum manifold. This naturally raises the question:
 Could dark space-time provide a physical medium that explains quantum non-locality and potentially allow faster-than-light (FTL) communication? In standard quantum mechanics, Bell’s theorem states that quantum mechanics is non-local but does not allow signal communication between entangled particles. Also, the no-signaling theorem ensures that measuring one particle of an entangled pair does not transmit information faster than light.
However, entangled particles instantaneously influence each other, which is still a mystery in standard physics. We still do not know what is the origin of the spooky infulence.  Here, we propose a solution by introducing dark space-time as an underlying structure that mediates quantum correlations in a hidden but fundamental way.

It must be noted that the experimental results have provided compelling evidence suggesting that quantum correlations, as observed in entangled particle pairs, exhibit influences that propagate faster than the speed of light—if one assumes any influence at all. The Gisin group \cite{scar,gisin} conducted long-distance Bell-type experiments using fast random setting changes and concluded that if any hidden communication explains the observed quantum correlations, it must travel at at least $10^4$ times the speed of light. Similarly, the Pan group \cite{pan}, demonstrated entanglement-based correlations violating Bell inequalities across vast distances and the speed of spooky action can be $10^4$ times higher than the speed of light. These results rule out any local hidden variable theories and suggest that, if quantum influences exist, they operate in a realm that transcends relativistic causal limits—highlighting the fundamentally nonlocal nature of quantum mechanics.

The main observation is that quantum entanglement in dark space-time may provide a hidden space-time connection.
This suggests that quantum entanglement occurs through a shared connection in dark space-time, rather than through standard space-time.
Each particle exists in both ordinary space-time and dark space-time.
When two particles become entangled, their dark space-time wavefunctions also become correlated.
This implies that when measurement is performed in ordinary space-time that collapses the wavefunction through dark space-time, enforcing non-locality without classical signals. This is similar in spirit to the scenario where a maximally entangled pair in ordinary quantum theory behaves differently \cite{akp} if subject to local unitary in the PT-symmetric quantum theory, and even can have signalling \cite{lee}. That is to say that local operation in a different world, can lead to faster than light communication in another world.

This suggests that the ``spooky action'' propagates in dark space-time rather than in ordinary space-time. 
This is also the ``speed of quantum information'' which moves at a superluminal
speed at which this ``influence'' should propagate from one station to the other one
limit to the speed of quantum information. Therefore, we may regard the dark space-time as the hidden layer of space-time for natural explanation of quantum nonlocality.
Dark space-time provides a causal structure for quantum entanglement, explaining its nonlocal nature.
All entangled states evolve jointly in both space-times, allowing superluminal effects in hidden dimensions.
This predicts new quantum anomalies that can be tested experimentally.

The experiments which show that quantum correlations between entangled particles cannot be explained by any local or slower-than-light influence—do indeed resonate conceptually with the notion of dark space-time. This hypothesis suggests that quantum nonlocality could be manifestations of an underlying invisible layer of space-time that is not accessible through classical means, but which could allow for faster-than-light influence. While these experiments do not directly prove the existence of dark space-time (as they remain agnostic about the mechanism behind quantum correlations), they align philosophically and physically with the idea that quantum mechanics involves deeper, nonlocal structures not captured by conventional relativistic space-time. Thus, the results provide indirect experimental support for the notion of dark space-time, which posit the existence of a deeper layer structure to explain quantum nonlocality and its possible cosmological connections. Future experiments that probe entanglement under varying gravitational or cosmological conditions may offer more direct evidence in support of such bold and unifying ideas.

\section{No-Hiding Theorem and Dark space-Time}
The No-Hiding Theorem \cite{pati} states that if quantum information is missing from one part of the system, it must be fully present in another subsystem. This prohibits quantum information from being lost in any classical or thermal environment—ensuring that information is never truly destroyed but only redistributed in a closed quantum system.
This theorem rules out total information loss, reinforcing the principle of conservation of quantum information in quantum mechanics.
The no-hiding theorem has been experimentally tested \cite{anil}.

If Dark Space-Time (DST) exists, it provides an additional hidden sector that interacts with quantum systems. 
The crucial question is can information be ``hidden'' within DST instead of being destroyed?
One can examine this by considering the evolution of quantum information in an extended Hilbert space that includes DST.
We now assume that quantum evolution happens in a two-space-time sector, i.e., ordinary ${\cal M}_{O}$
ordinary observable space-time and a dark space-time ${\cal M}_{D}$--a hidden dark space-time sector.
The quantum evolution happens in the total Hilbert space 
${\cal H}_{Total} = {\cal H}_{Ordinary} \otimes {\cal H}_{Dark} $.
If an initial state $|\Psi \rangle$  undergoes an evolution that ``hides" its information via an extended unitary process, then by 
the no-hiding Theorem, we will have conservation of Information in DST.
If the original quantum information is lost from the observable sector ${\cal H}_{Ordinary}$
Then it must be fully encoded in the hidden DST sector ${\cal H}_{Dark}$.
This proves that quantum information is never lost, rather it is simply transferred to DST.

The black hole information loss paradox \cite{haw} arises from the fact that Hawking radiation appears thermal and uncorrelated with the initial state of the black hole. If a black hole evaporates completely, standard physics suggests that information is lost, violating quantum unitarity. The presence of dark space-time provides an additional subsystem where quantum information can be transferred during black hole evaporation. Instead of being destroyed, information about the initial state of the black hole is stored in the DST sector via interactions with the event horizon. 

The Hawking radiation appears as a partial projection of a larger quantum state. Thus, the Hawking radiation is not purely thermal, but rather a projection of an entangled state that spans both ordinary space-time and dark space-time.
The hidden quantum correlations within dark space-time ensure that the total quantum system remains pure, preserving unitarity.
If dark space-time couples back to ordinary space-time at late times, then information previously hidden in DST can re-emerge in ordinary space-time, resolving the apparent loss.
This provides a natural mechanism for the Page curve \cite{page} which describes how information returns in black hole evaporation.
The Quantum Theory of Dark Space-Time (QDST) introduces a mathematical framework for modeling interactions between observed and hidden space-time. This shows how dark space-time acts as an information reservoir, resolving the black hole information paradox and may also explain dark energy as an entanglement-induced effect.

\section{Dark Space-Time and ER=EPR}
 
The ER=EPR conjecture, proposed by Maldacena and Susskind \cite{mald}, suggests a deep equivalence between quantum entanglement (EPR correlations) and Einstein-Rosen (ER) bridges, or wormholes. This conjecture implies that entangled particles are connected by nontraversable wormholes, offering a potential geometric explanation for quantum nonlocality.

The Dark Space-Time (DST) hypothesis, which postulates an additional hidden space-time structure, provides a natural setting to extend, support, and refine ER=EPR by offering a mechanism for the emergence of these wormhole connections through an extended space-time framework. Below, we explore this in detail.

The ER=EPR conjecture posits that Quantum Entanglement (EPR) $\equiv$ Geometric Connection via Wormholes (ER) \cite{rosen}. Thus, an entangled pair of particles, say Alice’s and Bob’s qubits, are not just mathematically correlated but are also physically connected via an underlying geometric structure in space-time. However, in standard relativity, wormholes are nontraversable, meaning they do not allow superluminal communication. Then a key question arises: how to explain entanglement between any particle and what space-time hosts these structures? In standard physics, a wormhole would exist within our known space-time, but its formation requires exotic matter, negative energy, or violations of energy conditions. 

Now, the notion of dark space-time can provide an alternative, i.e.,  such exotic bridges exist in dark space-time rather than ordinary space-time. The ER bridges do not need to exist in our space-time $M_{oridnary}$ but rather in an extended hidden sector 
dark $M_{dark} $. We speculate that particles entangled in ordinary space-time are geometrically connected in 
dark space-time by a traversable wormhole. Therefore, dark space-time offers a natural framework for non-local connectivity.
If the geometry of dark space-time differs from 
ordinary space-time, then two widely separated particles in ordinary space-time may be very close in dark space-time.
The effective distance in dark space-time (A,B), could be much smaller than the Minkowski separation (A,B), allowing the ER bridge to be a fundamental connection in dark space-time rather than in ordinary space-time.
If ER=EPR is true, then quantum information should be encoded in a way that maintains unitarity.
Therefore, dark space-time may provide a mechanism to store and transmit information through a hidden, conserved quantum structure, ensuring no information is lost when particles fall into a black hole.

One major application of ER=EPR is resolving the black hole information paradox. When matter falls into a black hole, its information should either be lost (as suggested by Hawking) or somehow recovered (to preserve unitarity of quantum theory).
According to ER=EPR, the Hawking radiation emitted from a black hole is entangled with the interior degrees of freedom.
If this entanglement is associated with an ER bridge, then information about the black hole interior is encoded in dark space-time rather than lost.
Information is stored in the dark space-time structure. The wormhole in dark space-time is not just a geometric curiosity but an information-preserving bridge. While information inside a black hole appears lost in ordinary space-time, it is actually preserved in dark space-time.
Hawking radiation emerges as a quantum process involving the dark space-time sector. The black hole horizon behaves like a quantum channel where information is transferred via dark space-time-mediated interactions.

If entanglement is mediated via dark space-time, then the Bell-type experiments at very large distances (e.g., cosmic-scale Bell tests) may show deviations from standard quantum predictions. Black holes may not completely evaporate but instead leave behind dark space-time remnants that store quantum information, preventing true information loss.
ER=EPR suggests that space-time itself emerges from entanglement.
The notion of dark space-time provides a natural background space-time for this emergence, offering a geometric foundation for quantum gravity.

Thus, dark space-time provides the geometric space in which ER=EPR bridges are physically realized, resolving locality issues.
Thus, the dark space-time hypothesis offers a robust extension of ER=EPR, providing a deeper understanding of both quantum nonlocality and black hole information physics. 

Now we address the question can dark space-time prove ER=EPR conjecture? 
This is a question that touches the heart of some of the deepest mysteries in modern theoretical physics. 
The concept of dark space-time, as proposed here, envisions a nonlocal, hidden substratum of reality that supports instantaneous redistribution of quantum information, thereby resolving the tension between quantum nonlocality and relativistic locality. This substratum lies beneath the visible geometry of conventional space-time and allows for non-spatiotemporal connections between entangled systems.

In ER=EPR, the bridge between two entangled particles is geometric. In dark space-time, the bridge is informational—yet real. If we posit that entanglement involves connections in a deeper layer of space-time (i.e., dark space-time), then these informational bridges could manifest as geometric wormholes when viewed from a higher-dimensional or dual description. Thus, dark space-time might provide the physical substrate that underlies the ER=EPR duality. One of the main issues in connecting quantum entanglement with wormholes is the preservation of causality. ER=EPR conjecture ensures this by making wormholes non-traversable. Similarly, dark space-time allows for nonlocal information redistribution without enabling superluminal signalling. Therefore, both frameworks preserve causality while allowing deep, nonlocal connections.

In some approaches to quantum gravity, space-time geometry itself is emergent from entanglement structure. Dark space-time could serve as the pre-geometric substrate, and ER=EPR could be seen as a macroscopic geometric manifestation of the microscopic entanglement structure encoded in dark space-time. However, a formal “proof” of ER=EPR via dark space-time
needs a precise mathematical formulation of dark space-time and demonstration that entangled states always correspond to traversals or mappings in this substratum. Also, we need to derive the space-time geometry (like ER bridges) as emergent structures from these mappings. In future, it will be worth developing a theory that can derive Einstein-Rosen bridges as emergent from quantum correlations via dark space-time, then dark
space-time could provide the foundational proof or mechanism behind ER=EPR.

While the ER=EPR conjecture arises in string theory and holography, and dark space-time is a concept emerging from quantum foundations, they may be two sides of the same coin. The dark space-time idea could provide a more general, information-theoretic framework within which ER=EPR becomes a special case—thus offering a path to a deeper unification of quantum mechanics and gravity.

\section{Speed in dark space-time}

 In ordinary space-time (Minkowski space), the speed of light 
$c$ sets the causal structure. Signals can only travel at or below $c$, ensuring no information moves faster than light. 
In the dark space-time (DST), we introduce a new fundamental speed $c_{dark} \gg c$, which governs how quickly quantum information propagates in DST.
If $c_{dark}$ is finite but large, information can propagate extremely fast in DST, much faster than in ordinary space-time.
If $c_{dark} \rightarrow \infty $ information can traverse any distance in DST in zero proper time. Further, 
the effective speed limit in DST diverges. The metric becomes singular, implying that the ``distance" between two points in DST shrinks to zero in proper time.
Any two points in DST become causally connected instantaneously (from the perspective of ordinary space-time).

Now consider an observer in ordinary space-time (Minkowski). If two entangled particles are linked via DST geodesics, then 
The correlation propagates through DST at speed $c_{dark}$.
If $c_{dark} \rightarrow \infty $  the propagation time in DST is zero.
From the viewpoint of an ordinary space-time observer, this would look like instantaneous action-at-a-distance.
This mechanism can explain quantum nonlocality without true faster-than-light (FTL) communication in ordinary space-time. Instead, the "instantaneous" effect is due to the fact that in DST, the two points are already connected with effectively zero separation.
The intuition behind  the ``shortcut" effect is in an analogy with the wormhole shortcut idea in general relativity (ER=EPR conjecture).
In ordinary space-time, two locations A and B are separated by a large distance.
In DST, these two locations may be very close or even connected directly.
A signal traveling in DST would appear instantaneous in ordinary space-time, just like a wormhole allows a shortcut between distant points.

One may ask does this Violate Causality? The answer is no, because no information travels faster than light in ordinary space-time.
The effect is purely due to dark space-time geometry, which is separate from the Minkowski causality.
The notion of dark space-time provides a hidden variable-like space where correlations exist without direct faster than light communication.
Thus, entanglement correlations could be understood as an effect in dark space-time which appear nonlocal in ordinary space-time.

\section{Quantum Teleportation and Dark Space-Time}
In standard quantum mechanics, an unknown quantum state to be teleported is given to Alice.
Alice and Bob share a maximally entangled pair. Alice performs a Bell-state measurement on her half of the entangled pair and the unknown state.
She sends 2 classical bits to Bob. Next, Bob, using those 2 bits, performs a local unitary operation on his qubit to recover the original quantum state \cite{cb}.

However, quantum teleportation does raise an apparent puzzle: how does the complete quantum state which contains potentially an infinite amount of bits get transferred from Alice to Bob using only 2 classical bits?
The actual quantum state contains uncountably many amplitudes and phase information—so where does this quantum information go?
It just happens in a magical way. 
Can dark space-time throw some light? Can dark space-time explain how information flows from Alice to Bob in quantum teleportation?

Note that no actual “infinite bits of information” is transmitted through the classical channel. The entanglement acts like a quantum resource that, along with the classical bits, enables the reconstruction of the state. But standard QM says no information travels faster than light as the 2 bits are necessary and sufficient to complete the process. 
Jozsa has suggested that the potentially vast quantum information must have been propagated
backwards in time from Alice to the EPR source and thence forwards in time to Bob \cite{rj}.

In the dark space-time framework, we hypothesize that quantum information flows through a hidden layer of space-time, one that is not directly observable, yet mediates entanglement-based correlations in real time—possibly instantaneously or at superluminal speeds.

In this view, we can say that the entangled pair is connected via a dark space-time wormhole-like bridge (a "quantum thread").
When Alice performs her measurement, the quantum information about the unknown state is instantly updated at Bob’s location via this hidden layer. The 2 classical bits merely tell Bob which local operation to apply to recover the quantum state.
Thus, the "infinite amount of of information" in the quantum state does not travel through classical space-time.
Instead, it is re-routed through the dark space-time, ensuring that Bob’s qubit acquires the right state before or as soon as the classical bits arrive.

Quantum teleportation challenges our understanding of information flow. Dark space-time may offer a compelling hypothesis: the information does not flow through space-time at all in the conventional sense. It flows through a hidden quantum layer, a deeper structure of the universe. This could explain how the full structure of a quantum state is “transferred” with only two classical bits.

\section{New Predictions of the Existence of Dark Space-Time}

The Quantum Theory of Dark Space-Time (QDST) provides a new framework that extends standard quantum gravity, black hole physics, and cosmology. Below are key testable predictions that could provide evidence for its existence.

1. {\it Gravitational Decoherence and Modified Quantum Experiments}: If quantum systems interact with dark space-time, we should observe gravitationally-induced decoherence beyond standard predictions from environmental noise.
This may be observed in interferometry with macroscopic quantum systems.
In standard quantum mechanics, decoherence occurs due to the interaction with known environments (e.g., air molecules, photons).
QDST predicts an additional universal decoherence effect due to information leakage into dark space-time.
If we can conduct massive interferometer experiments (e.g., levitated nanoparticles, Bose-Einstein condensates) and check if coherence loss follows the predicted scaling law:
\begin{align}
    \Gamma_{dark} = \frac{m^2}{M_{P}^2} e^{-\alpha t},
\end{align}
where $\Gamma_{dark}$ is the rate of decoherence due to dark space-time interaction, $M_P$ is the Planck mass and $\alpha$ is a constant. If unexplained decoherence is observed in interference experiment beyond known sources, it could signal the presence of dark space-time interactions.

2. {\it  Modified Page Curve for Black Hole Evaporation}: If dark space-time exists, then black hole evaporation will not follow the standard Page curve but will saturate more slowly, as quantum information leaks into dark space-time.
One possible test can be to have simulation of black hole evaporation in quantum simulators.
Analog gravity experiments \cite{ana} (e.g., Hawking radiation in Bose-Einstein condensates, optical analogs) can test deviations from standard semiclassical behavior. A delayed saturation of the entanglement entropy of Hawking radiation due to information leakage into dark space-time can have observable implication.
If experimental quantum black hole models show anomalous information recovery rates, it would support the idea that hidden degrees of freedom in dark space-time store information.

3.{\it Gravitational Wave Signatures from Dark Space-Time Coupling}:
If dark space-time exists, then gravitational waves should show anomalous dissipation or additional polarization modes due to coupling between ordinary and dark space-time. Experimental 
observation of gravitational waves \cite{ligo} with LIGO, LISA, or next-generation detectors may 
 suggest an extra damping term in the propagation of gravitational waves, due to energy leakage into dark space-time.
The wave equation gets an additional dissipation term due to the leakage in the dark space-time sector.
Hence, if we look for deviations from the General Relativity predictions in the decay rates of gravitational waves such as the 
anomalous wave attenuation in gravitational wave signals, then that could be a direct test of energy exchange with dark space-time.

4. {\it Dynamical Dark Energy and Evolution of the Cosmological Constant}:
There is evidence for the evolution of the cosmological constant on the cosmological scale \cite{zla}.
If dark space-time exists, then it may provide another explanation, i.e., the effective cosmological constant is not truly constant but may evolve due to the interaction with dark space-time. This could explain why the universe’s accelerated expansion is recent. 
Precise measurements of late-time cosmic expansion in experiments may indicate 
if dark space-time contributes dynamically to dark energy. If so, we should see small deviations from 
the current predictions, particularly in the Hubble parameter’s evolution.
Next-generation telescopes could detect small deviations in supernova redshift-luminosity relations and Baryon Acoustic Oscillations.
If observational data reveals a slow decay in dark energy density, it would suggest an underlying effect involving dark space-time.\\

\section{Conclusion}

In this work, we have introduced the concept of dark space-time as a hidden geometric structure that coexists with ordinary space-time and provides a natural explanation for quantum nonlocality. By formulating a two-space-time framework, we have shown that entangled states can evolve through interactions between ordinary and dark space-time sectors, effectively mediating nonlocal correlations via an unobservable but causal channel. This approach challenges the conventional understanding of locality while maintaining mathematical consistency within a modified space-time metric.

Furthermore, we explored the implications of DST for the ER=EPR conjecture, suggesting that wormhole-like structures in dark space-time could serve as geometric bridges underlying quantum entanglement. We also examined the potential role of dark space-time in resolving the black hole information paradox, where hidden degrees of freedom in the dark sector may store and later release information, preserving unitarity. Importantly, our analysis indicates that dark space-time is a more fundamental entity than dark matter and dark energy, potentially providing a unified geometric foundation for these mysterious cosmological components.

Future research should focus on deriving observable consequences of dark space-time, including potential deviations from General Relativity, modifications to the cosmic microwave background, and new experimental signatures in quantum information processing. 
A direct test of dark space-time would be as revolutionary as the Michelson–Morley experiment was for Einstein’s relativity. It requires bold theoretical proposals and cutting-edge experimental precision.
If validated, this framework could represent a paradigm shift in our understanding of space-time, causality, and the quantum fabric of reality.


\begin{thebibliography}{0}%
\makeatletter
\providecommand \@ifxundefined [1]{%
 \@ifx{#1\undefined}
}%
\providecommand \@ifnum [1]{%
 \ifnum #1\expandafter \@firstoftwo
 \else \expandafter \@secondoftwo
 \fi
}%
\providecommand \@ifx [1]{%
 \ifx #1\expandafter \@firstoftwo
 \else \expandafter \@secondoftwo
 \fi
}%
\providecommand \natexlab [1]{#1}%
\providecommand \enquote  [1]{``#1''}%
\providecommand \bibnamefont  [1]{#1}%
\providecommand \bibfnamefont [1]{#1}%
\providecommand \citenamefont [1]{#1}%
\providecommand \href@noop [0]{\@secondoftwo}%
\providecommand \href [0]{\begingroup \@sanitize@url \@href}%
\providecommand \@href[1]{\@@startlink{#1}\@@href}%
\providecommand \@@href[1]{\endgroup#1\@@endlink}%
\providecommand \@sanitize@url [0]{\catcode `\\12\catcode `\$12\catcode `\&12\catcode `\#12\catcode `\^12\catcode `\_12\catcode `\%12\relax}%
\providecommand \@@startlink[1]{}%
\providecommand \@@endlink[0]{}%
\providecommand \url  [0]{\begingroup\@sanitize@url \@url }%
\providecommand \@url [1]{\endgroup\@href {#1}{\urlprefix }}%
\providecommand \urlprefix  [0]{URL }%
\providecommand \Eprint [0]{\href }%
\providecommand \doibase [0]{http://dx.doi.org/}%
\providecommand \selectlanguage [0]{\@gobble}%
\providecommand \bibinfo  [0]{\@secondoftwo}%
\providecommand \bibfield  [0]{\@secondoftwo}%
\providecommand \translation [1]{[#1]}%
\providecommand \BibitemOpen [0]{}%
\providecommand \bibitemStop [0]{}%
\providecommand \bibitemNoStop [0]{.\EOS\space}%
\providecommand \EOS [0]{\spacefactor3000\relax}%
\providecommand \BibitemShut  [1]{\csname bibitem#1\endcsname}%
\let\auto@bib@innerbib\@empty
\end{thebibliography}%


\begin{thebibliography}{99}

\bibitem{cow} 
 R. Cowen, 
 Nature {\bf 527} 290 (2015)

\bibitem{miao} L. Miao, L. Xiao-Dong, W. Shuang, Wang Yi, Frontiers of Physics, {\bf 8}, 828 (2013).


\bibitem{ashte} 
A. Ashtekar, E. Bianchi
Rep. Prog. Phys. {\bf 84}, 042001 (2021).

\bibitem{epr} 
A. Einstein, B. Podolsky and N. Rosen, 
Physical Review {\bf 47}, 777 (1935).

\bibitem{bell} J. S. Bell, Physics (Long Island City, N.Y.) {\bf 1}, 195 (1964). 

\bibitem{chsh} J. F. Clauser, M. A. Home, A. Shimony, and R. A. Holt, Phys. Rev. Lett. {\bf 23}, 880 (1969).

\bibitem{mald} J.  Maldacena and L. Susskind, Leonard,
Fortschritte der Physik {\bf 61}, 781  (2013).

\bibitem{rosen}
A. Einstein and N. Rosen,
Physical Review {\bf 48}, 73 (1935).


\bibitem{van} 
M. van Raamsdonk, 
International Journal of Modern Physics D {\bf 42},  2323 (2010).

\bibitem{am}
A. Arbey and F. Mahmoudi,
Progress in Particle and Nuclear Physics {\bf 119},  103865 (2021).


\bibitem{scar} V. Scarani, W. Tittel, H. Zbinden, and N. Gisin, Phys. Lett. A {\bf 276}, 1 (2000).

\bibitem{gisin}  D. Salart, A. Baas, C. Branciard, N. Gisin, and H. Zbinden, Nature 454, 861 (2008).

\bibitem{pan} Juan Yin {\it et al}, Phys. Rev. Lett. {\bf 110}, 260407 (2013).


\bibitem{akp} A. K. Pati, Pramana-J. of Physics {\bf 73}, 485 (2009).

\bibitem{lee} Yi-Chan Lee, Min-Hsiu Hsieh, Steven T. Flammia, and Ray-Kuang Lee, Phys. Rev. Lett. {\bf 112}, 130404 (2014).


\bibitem{pati} S. L. Braunstein and A. K. Pati, Phys. Rev. Lett. {\bf 98}, 080502 (2007).

\bibitem{anil} J. R. Samal, A. K. Pati and Anil Kumar,  Phys. Rev. Lett. {\bf 106} 080401 (2011).

\bibitem{haw} S. W. Hawking, Phys. Rev. D. {\bf 14}, 2460 (1976).

\bibitem{page} D. Page, 
Phys. Rev. Lett. {\bf 71} 3743 (1993).

\bibitem{cb}  C. Bennett, G. Brassard, C. Crepeau, R. Jozsa, A. Peres,
W.K. Wooters, Phys. Rev. Lett. {\bf 70} 1895 (1993).

\bibitem{rj} R. Jozsa,
IBM Journal of Research and Development, {\bf 48}, 79 (2004).

\bibitem{ana} C. Barceló, S. Liberati, M. Visser, 
Classical and Quantum Gravity. {\bf 18}, 1137 (2001).

\bibitem{ligo} P. B. Abbott, {\it  et al.} 
Phys. Rev. Lett. {\bf 116}, 061102 (2016).

\bibitem{zla} I. Zlatev, L. Wang, P. J. Steinhardt, Phys. Rev. Lett. {\bf 82}, 896 (1999).

\end{thebibliography}
\end{document}